\documentclass[12pt]{iopart}

\usepackage{graphicx}   
\usepackage{longtable}
\usepackage{lineno,hyperref}
\usepackage{iopams}
\usepackage{cite}

\usepackage{epstopdf} 
\begin{document}

	\title[SWKB is not exact for all SIPs]{The Supersymmetric WKB Formalism is Not Exact for All Additive Shape Invariant Potentials}

\author{Jonathan Bougie, Asim Gangopadhyaya, Constantin Rasinariu}
\eads{\mailto{jbougie@luc.edu}, \mailto{agangop@luc.edu}, \mailto{crasinariu@luc.edu}}

\address{Department of Physics, Loyola University Chicago, Chicago, IL 60660, U.S.A.}

%
%
%
%
		
\date{April 10, 2018}
	
\begin{abstract}
Following the verification of the conjecture made by Comtet, Bandrauk and Campbell that the supersymmetry-inspired semiclassical method known as SWKB is exact for the conventional additive shape invariant potentials, it was widely believed that SWKB yields exact results for all additive shape invariant potentials. In this paper we present a concrete example of an additive shape invariant potential for which the SWKB method fails to produce exact results.
\end{abstract}




\maketitle

\section{Introduction}

	
The well known JWKB method \cite{Jeffreys,Wentzel,Kramers,Brillouin} is
 a semiclassical approximation method
 used to generate solutions of the Schr\"odinger equation as a power series in  $\hbar$ \cite{Bender_Orzag}.
 The lowest order term in this approximation gives the following quantization condition for the energy

\begin{linenomath}
\begin{equation}
	\int_{x_L}^{x_R} \sqrt{E_n-V(x)}\quad {\rm d}x = \left( n+\frac12\right) 
	\pi\hbar~, \quad \mbox{where}~ n=0,1,2,\cdots \label{eq:jwkb}~~.
\end{equation}
\end{linenomath}
The integration limits $x_L$ and $x_R$ are the classical turning points on the $x$-axis given by ${E_n-V(x)}=0$. In several cases, this quantization condition produces exact spectra \cite{Bender}. 
	
In 1985, Comtet et al.\cite{Comtet}, in the context of supersymmetric quantum mechanics (SUSYQM), proposed  a variant of the above condition  and showed that it generated exact spectra for all known solvable systems at that time \cite{Infeld, Dutt_SUSY}. This modified quantization condition, known as the Supersymmetric WKB or SWKB, prescribes 
\begin{linenomath}
\begin{equation}
	\int_{x_L}^{x_R} \sqrt{E_n-W^2(x)}\quad{\rm d}x = n \pi\hbar~, \quad
	\mbox{where}~ n=0,1,2,\cdots 
	\label{eq:swkb1}.
\end{equation}
\end{linenomath}
Here, $W(x)$ is the superpotential that generates a given potential by $ V(x) = -W'(x)+W^2(x)$. For the SWKB method, the integration limits $x_L$ and $x_R$ are the turning points on the $x$-axis given by ${E_n-W^2(x)}=0$. In a very interesting result \cite{Dutt}, Dutt et al. showed that the SWKB method, at the lowest order, generates correct spectra for an entire class of potentials known as conventional additive shape invariant potentials.  Additionally, it was shown\cite{Adhikari} that the higher order corrections to the eigenvalues vanish to ${\cal O}(\hbar^6)$ for all  conventional additive shape invariant potentials \cite{Infeld, Dutt_SUSY}. Subsequently,  Raghunathan et al. \cite{Raghunathan} argued that all higher order contributions vanish as well.
This prompted the conjecture \cite{Dutt, Adhikari, Raghunathan, Barclay, Yin} that, for reasons yet to be determined, SWKB yields exact eigenspectra for all additive shape invariant potentials. 
	
In this paper we present a concrete example of an additive shape invariant superpotential for which the SWKB method fails to produce exact results. This superpotential belongs to a class of additive shape invariant superpotentials known as extended superpotentials\cite{Quesne1,Quesne2,Odake1,Odake2}. Unlike the conventional shape invariant superpotentials, these have an inherent dependence on $\hbar$.
	
The paper is organized as follows. In the next section we provide a brief introduction to supersymmetric quantum mechanics (SUSYQM), shape invariance,  $\hbar$-dependent superpotentials, and SWKB.  In Section \ref{sec:results} we employ a combination of analytic and numeric methods to present a concrete example of an additive shape invariant potential for which the SWKB method is not exact.

\section{Preliminaries}
\subsection{SUSYQM}
The formalism of supersymmetric quantum mechanics  \cite{Witten,Solomonson,CooperFreedman}
can be viewed as a generalization of the Dirac-Fock oscillator method for solving the eigenvalue problem of a harmonic oscillator. In SUSYQM \cite{Cooper-Khare-Sukhatme, Gangopadhyaya-Mallow-Rasinariu}, one replaces the ladder operators of the harmonic oscillators $b^{\pm}= \mp\, \hbar\; d/dx+ 1/2 \;\omega x$ by
${\cal A}^\pm = \mp \, \hbar\; d/dx+ W(x,a)$, where $a$ is a parameter.  The function $ W(x,a)$ is known as the
superpotential. For simplicity, we have set the mass $2m = 1$. The product of operators ${\cal A}^+$ and ${\cal A}^-$ produces a hamiltonian $H_-={\cal A}^+\!\cdot {\cal A}^-$ given by
\begin{linenomath}
	\begin{eqnarray}
	{\cal A}^+\!\cdot {\cal A}^-
	\!\!&=&\! \left(  -\hbar \frac{d}{dx}+ W(x,a) \right)  
	~\left( \hbar \frac{d}{dx}+ W(x,a)\right) \nonumber \\
	\!\!&=&\!  -\hbar^2 \frac{d^2}{dx^2}+W^2(x,a)  -  \hbar \frac{dW}{dx}~ 
	\label{A+A-}~.
	\end{eqnarray}
\end{linenomath}
The corresponding potential $V_-(x,a)$ is related to the superpotential by $V_-(x,a) = W^2(x,a) - \hbar ~{d W}\!/{dx}$. Given the semi-positive definite nature of $H_-$, its eigenvalues $E^{-}_{n}$ are either positive or zero. If the lowest eigenvalue $E^{-}_{0}\neq 0$, the system is said to have broken supersymmetry. Henceforth we will consider systems with unbroken supersymmetry, i.e., their lowest eigenvalue is zero. 
	
The product ${\cal A}^-\!\cdot {\cal A}^+$ generates another
hamiltonian $H_+= -\hbar^2 \frac{d^2}{dx^2}+V_+(x,a)$ with $V_+(x,a) =
W^2(x,a) + \hbar \;d\, W/dx$. The two hamiltonians are related: 
${\cal A}^+\!\cdot H_+ = H_- \!\cdot {\cal A}^+$ and ${\cal A}^-\!\cdot H_-
	= H_+\!\cdot{\cal A}^-$.  
This intertwining leads to the following relationships among the eigenvalues and eigenfunctions of the partner hamiltonians $H_-$ and $H_+$: 
\begin{linenomath}
\begin{equation}
	E^{-}_{n+1} =E^{+}_{n}, \quad  n=0,1,2,\cdots~ 
\end{equation}
\begin{eqnarray}
\frac{~~~{\cal A}^- }{\sqrt{E^{+}_{n} }} ~\psi^{-}_{n+1} 
	= ~\psi^{+}_{n}  ~;\quad
	\frac{~~~{\cal A}^+}{\sqrt{E^{+}_{n} }}~\psi^{+}_{n} 
	= ~  \psi^{-}_{n+1}~. \label{isospectrality}
\end{eqnarray}
\end{linenomath}
Thus, if we knew the eigenvalues and eigenfunctions of either of the two partner hamiltonians, we could determine the eigenvalues and the eigenfunctions of the other. This property is known as  isospectrality. 
	
\subsection{Shape Invariance}
Let us consider a set of parameters $a_i\,,i=0,1,\cdots$. We choose $a_0=a$ and $a_{i+1} = f(a_i)$ where $f$ is an arbitrary function that models the parameter change. A superpotential $W(x,a_i)$ is called ``shape invariant'' (SI) if it  obeys the following  condition,
\begin{linenomath}
\begin{equation}
	W^2(x,a_i)  +  \hbar \frac{d\, W(x,a_i)}{dx}+g(a_i) = 
	W^2(x,a_{i+1})  -  \hbar \frac{d\, W(x,a_{i+1})}{dx}+g(a_{i+1})~,
\label{SIC1}
\end{equation}
\end{linenomath}
where $g$ is some function only of the parameters $a$.
The eigenvalues and eigenfunctions of shape invariant potentials are given by \cite{Infeld,Miller,gendenshtein1,gendenshtein2} 
\begin{linenomath}
\begin{eqnarray}
E_n^{-}(a_0) & = & g(a_n)-g(a_0)\\
\psi^{-}_{n}(x,a_0)& = &
\frac{{\cal A}^+{(a_0)} 
	~ {\cal A}^+{(a_1)}  \cdots  {\cal A}^+{(a_{n-1})}}
{\sqrt{\,E^{-}_n(a_0)\,E_{n-1}^{-}(a_1)\cdots E_{1}^{-}(a_{n-1})}}~\psi^{-}_0(x,a_n)
~,
\end{eqnarray}
\end{linenomath}
where the ground state $\psi^{-}_0(x,a_n) = N \; exp\left[-\frac1\hbar\, \int^x W(y,a_n)\,dy\right]$ is the solution of ${\cal A}^-\psi^{-}_0 =0 $.
	
In this paper we will consider the case of additive parameter change: $a_{i+1}=a_{i}+\hbar$. A complete list of additive SI superpotentials that do not explicitly depend on $\hbar$ was given in Refs. \cite{Infeld,Dutt_SUSY,CGK}. We call these  ``conventional superpotentials.'' 
	
In Refs. \cite{Quesne1, Quesne2, Odake1,Odake2, Tanaka, Odake3, Odake4, Quesne2012a, Quesne2012b, Ranjani1, Ranjani2} a new class of additive shape invariant superpotentials was found that depends  explicitly on $\hbar$. Such superpotentials are called ``extended." 
	
In Ref. \cite{Bougie2010, symmetry} the authors showed that there exists a special set of partial differential equations that describe both the conventional and extended superpotentials. The conventional superpotentials  are obtained by expanding $W(x,a_i+\hbar)$ in powers of $\hbar$, assuming that $W$ is independent of $\hbar$ except through the parameter $a$ and substituting back into (\ref{SIC1}). This equation must hold for an arbitrary value of $\hbar$. Thus the coefficient of each power of $\hbar$ must independently vanish. Hence, for various powers of $\hbar$ we get
\begin{linenomath}
\begin{eqnarray}%
W \, \frac{\partial W}{\partial a} - \frac{\partial W}{\partial x} + \frac12 \, \frac{d g(a)}{d a} = 0~&\quad\quad {\cal O}(\hbar) 
\label{PDE1a}\\
 \frac{\partial }{\partial a}\left( W \, \frac{\partial W}{\partial a} - \frac{\partial W}{\partial x} + \frac12 \, \frac{d g(a)}{d a} \right)= 0~&
\quad\quad 
{\cal O}(\hbar^2)
\label{PDE2}\\
 \frac{\partial^{n}}{\partial a^{n-1}\partial x} ~W(x,a)= 0~, ~~~~~n\geq 3 ~&\quad\quad 
{\cal O}(\hbar^n)\, .
\label{PDE3} \end{eqnarray}
\end{linenomath}
Although this represents an infinite set, if equations of ${\cal O}(\hbar)$ and ${\cal O}(\hbar^3)$ are satisfied, all others automatically follow.  Therefore, to find the complete set of solutions, it is sufficient to solve:
\begin{linenomath}
\begin{eqnarray} 
W \, \frac{\partial W}{\partial a} - \frac{\partial W}{\partial x} + \frac12 \, \frac{d g(a)}{d a} = 0~
\label{PDE1}
\end{eqnarray} and
\begin{eqnarray}
\frac{\partial^{3}}{\partial a^{2}\partial x} ~W(x,a)= 0\label{Eq1}~.
\end{eqnarray}
The extended superpotentials having an inherent dependence on $\hbar$, can  be expanded in terms of powers of $\hbar$:
\begin{eqnarray}
W(x, a, \hbar) = \sum_{j=0}^\infty \hbar^j W_j(x,a)~.~ \label{W-hbar}
\end{eqnarray}
This power series when substituted in (\ref{SIC1}) yields \\
for $j=1$
\begin{eqnarray}
2\frac{\partial W_{0}}{\partial x}
-\frac{\partial }{\partial a} \left(W_{0}^2+g  \right) = 0~,
\label{lowestorder}
\end{eqnarray}
\end{linenomath}
for $j = 2$
\begin{linenomath}
\begin{eqnarray}
\frac{\partial W_{1}}{\partial x}
-
\frac{\partial}{\partial a} \left( W_0\, W_{1}\right)  = 0~,
\label{secondorders}
\end{eqnarray}
\end{linenomath}
and for $j \geq 3$
\begin{linenomath}
\begin{eqnarray}
2\,\frac{\partial W_{j-1}}{\partial x}
-\sum_{s=1}^{j-1} \sum_{k=0}^s \frac{1}{(j-s)!}
\frac{\partial^{j-s}}{\partial a^{j-s}} W_k\, W_{s-k}+
\nonumber \\
\sum_{k=2}^{j-1} \frac{1}{(k-1)!}
\frac{\partial^{k}\, W_{j-k}}{\partial a^{k-1} \,\partial x} 
+ \left( \frac{j-2}{j!}\right)\,
\frac{\partial^{j} W_0}{\partial a^{j-1}\partial x}   
=0~.
\label{higherorders}
\end{eqnarray}
\end{linenomath}
We first observe that (\ref{lowestorder}) is equivalent to
(\ref{PDE1}); i.e., all conventional shape invariant potentials
automatically satisfy (\ref{lowestorder}).  Hence, conventional
superpotentials can be used as the base to erect the tower of extended superpotentials. If
$W_0$ is a conventional superpotential, then owing to (\ref{Eq1}), the last term of (\ref{higherorders}) disappears.

For example, a superpotential that appears in\cite{Quesne1} can be built upon the conventional superpotential for the 3-D oscillator, $W_{0} = 1/2 \;\omega x -  \ell /x$, by setting $W_{1}=0$. The higher order terms of the superpotential $W$ can be generated from (\ref{higherorders}) for all $j>1$. We obtain
\begin{eqnarray}
	W(x,\ell) &=& \frac{\omega x}2  - \frac {\ell}x + \left(\frac{2\omega x \hbar}{\omega x^2 + 2\ell -\hbar}-\frac{2 \omega x \hbar}{\omega x^2+2\ell+\hbar}\right)~, \label{ExtendedSuperpotential-Quesne}
\end{eqnarray}
which is an $\hbar$-dependent additively shape invariant superpotential.
	
\subsection{SWKB}
As we saw in  (\ref{A+A-}), the potential in SUSYQM formalism is given by $W^2(x,a)-\hbar~ W'$, where $W'\equiv {dW}\!/{dx}$. Comtet et al. \cite{Comtet} interpreted the second term, due to the presence of $\hbar$, as a term generated by integration over fermionic degrees of freedom, and hence inherently quantum mechanical in nature. Consequently, they dropped this $\hbar$-dependent term from the potential for the semiclassical condition given in  (\ref{eq:jwkb}). This reduces the integral in (\ref{eq:jwkb}) to $ \int_{x_L}^{x_R} {\rm d}x\,\left[  E_n-W^2(x,a)\right]^{1/2}$, where this time,   $\left({x_L},{x_R} \right) $ are the turning points defined by $E_n-W^2(x,a) =0$. Furthermore, for the RHS of (\ref{eq:jwkb}), Comtet et al claimed that $\left( n+1/2\right) \pi\hbar$ should be replaced by $n \pi\hbar$, which was supported in Ref. \cite{Dutt} by expanding the LHS of (\ref{eq:jwkb}) in powers of $\hbar$. Thus, the semiclassical quantization condition in SUSYQM reduces to the following SWKB condition:
\begin{linenomath}
	\begin{equation}
	\int_{x_1}^{x_2} \sqrt{E_n-W^2(x,a)}\quad{\rm d}x = n \pi\hbar~, \quad \mbox{where}~ n=0,1,2,\cdots \label{eq:swkb}.
	\end{equation}
\end{linenomath}
Note that this was supposed to be an approximate condition. The attractiveness of SWKB comes from the fact that it generates correct spectra for all conventional additive shape invariant potentials \cite{Dutt}. This unexpected property led to the conjecture that in fact, SWKB is exact for all additive shape invariant potentials. In the next section we prove that this is not the case, by providing a concrete counterexample. We show that for an extended superpotential, shape invariance is, in fact, not a sufficient condition for SWKB-exactness. This is in agreement with recent speculation \cite{mahdi2016} that the $\hbar$-dependent terms in extended superpotentials could break the exactness of SWKB, and indicates that further study of the $\hbar$-dependence of extended superpotentials could be helpful for increasing our understanding of these superpotentials. 

\section{Non-Exactness of SWKB for an Extended Shape-Invariant Superpotential}\label{sec:results}

\subsection{Extended Radial Oscillator as an Example Superpotential}

We now show that additive shape invariance is not a sufficient condition for the exactness of the SWKB approximation by using the superpotential of the extended radial oscillator given in \cite{Quesne1} as a specific counter-example. This superpotential, given in (\ref{ExtendedSuperpotential-Quesne}), can be written as\footnote{This is equivalent to the superpotential of reference \cite{Quesne1}, with the identification $\ell \rightarrow \ell+1$ and $\hbar=1$.}
\begin{linenomath}
\begin{equation}
W=W_0+W_{h}~,
\label{9}
\end{equation} 
\end{linenomath}
where  
\begin{linenomath}
\begin{equation} 
W_0=\frac{1}{2}\omega x-\frac{\ell}{x}~,
\label{eq:W0}
\end{equation}
\end{linenomath}
and 
\begin{linenomath}
\begin{equation}
W_{h}=\left(\frac{2\omega x \hbar}{\omega x^2 + 2\ell -\hbar}-\frac{2 \omega x \hbar}{\omega x^2+2\ell+\hbar}\right)~.\label{eq:hbar1W}
\end{equation}
\end{linenomath}
Henceforth we set $\hbar=1$.

To simplify the problem, we now show that the SWKB approximation does not depend on $\omega$ for this superpotential. We begin with the superpotential given by Eqs.(\ref{9}-\ref{eq:hbar1W})
\begin{equation}
W\left(x,\ell\right)=\frac{1}{2}\omega x-\frac{\ell}{x}+\left(\frac{2\omega x}{\omega x^2 + 2\ell - 1}-\frac{2 \omega x }{\omega x^2+2\ell+1}\right)~,
\label{12}
\end{equation}
for which the SWKB approximation is given by 
\begin{equation}
\int_{x_L}^{x_R} \sqrt{E_n-W^2(x,\ell)}\quad{\rm d}x = n \pi~, \quad \mbox{where}~ n=0,1,2,\cdots.
\end{equation}
The two turning points $x_L$ and $x_R$ are given by positive solutions of 
\begin{equation}
E_n-W^2(x,\ell)=0~,
\end{equation}
where $E_n=2 n \omega$.

With the change of variable $ y\equiv\sqrt{\omega}x$, we obtain
\begin{equation}
W\left(y,\ell\right)=\sqrt{\omega}\left[\frac{1}{2}y-\frac{\ell}{y}+\left(\frac{2y}{y^2 + 2\ell - 1}-\frac{2 y }{y^2+2\ell+1}\right)\right]~.
\label{fullW}
\end{equation}
The SWKB approximation is given by 
\begin{equation}
\int_{x_L}^{x_R} \sqrt{E_n-W^2(x,\ell)}\quad{\rm d}x = \int_{y_L}^{y_R}\sqrt{\eta\left(y,\ell\right)}\quad{\rm d}y=n\pi\hbar~,
\end{equation}
where $\eta\left(y,\ell\right)=2n-\left(\frac{1}{2}y-\frac{\ell}{y}+\frac{4y}{\left(y^2 + 2\ell - 1\right)\left(y^2+2\ell+1\right)}\right)^2,$ and the integration limits are given by the zeros of $\eta\left(y_L,\ell\right)=\eta\left(y_R,\ell\right)=0.$ Note since $\eta\left(y,\ell\right)$ is independent of $\omega$, the SWKB approximation also does not depend on $\omega$. Therefore, without loss of generality, for the remainder of this paper we return to our original nomenclature by re-naming $y$ back to $x$ in (\ref{fullW}), which is equivalent to setting $\omega=1$ in (\ref{12}). 

We know that SWKB is exact for $W_0$, so we investigate whether adding $W_h$  changes this exactness.  We do this by perturbing the conventional potential via a parameter $\alpha$, such that 
\begin{equation} 
	W(x,\ell,\alpha )=W_0+\alpha W_h ~.
\end{equation}
We also parameterize the integral given by the SWKB approximation: 
\begin{equation}
I(n,\ell,\alpha ) = \int_{x_L}^{x_R} \sqrt{E_n-W^2(x,\ell,\alpha )}\quad{\rm d}x~, \label{Integral_alpha}
\end{equation}
where the integration limits $x_L$ and $x_R$ are solutions of 
$E_n-W^2(x,\ell,\alpha)=0$.

For the conventional case  ($\alpha=0$) the lowest order SWKB yields exact result, i.e., 
$$
I(n,\ell,0)= n \pi~, \quad \mbox{where}~ n=0,1,2,\cdots~.
$$
In the following sections, we investigate how the SWKB integral changes as $\alpha$ is increased from zero to one, beginning with small perturbations from the exact result at $\alpha=0$. Note that the superpotential is shape invariant for the conventional case $W(x,\ell,0)$ and for the extended case $W(x,\ell,1)$, but not for a general value of $\alpha$.

\subsection{$\alpha$--Dependence of the SWKB approximation near $\alpha=0$}\label{smallalpha}

We first examine the behavior of the SWKB approximation near $\alpha=0$. From (\ref{Integral_alpha}), we have
\begin{eqnarray}
	\frac{\partial I}{\partial \alpha} &=&
	\left. \frac{\partial {x_R}}{\partial \alpha} \sqrt{E_n-W^2(x,\ell,\alpha )}\right|_{x=x_R}
	-
	\left. \frac{\partial {x_L}}{\partial \alpha} \sqrt{E_n-W^2(x,\ell,\alpha )}\right|_{x=x_L}
	\nonumber \\ 
	&+& 
	\int_{x_L}^{x_R} \frac{\partial }{\partial \alpha}  \sqrt{E_n-W^2(x,\ell,\alpha )}\quad{\rm d}x.  \label{Variation_of_limits_term} 
\end{eqnarray}
The first two terms on the RHS in (\ref{Variation_of_limits_term}) drop out as the factors under the radical signs vanish at the turning points. Hence, 
\begin{eqnarray}
	\frac{\partial I}{\partial \alpha}  &=& \int_{x_L}^{x_R} \frac{\partial }{\partial \alpha}  \sqrt{E_n-\left(W_0 + \alpha W_h \right)^2}\quad{\rm d}x
	\nonumber \\
	&=& - \int_{x_L}^{x_R}  \frac{ W_0 W_h + \alpha W_h^2   }{\sqrt{E_n-W_0^2-2\alpha W_0 W_h -\alpha^2 W_h^2 }}\quad{\rm d}x ~.
\end{eqnarray}

As $\alpha$ goes to zero, this yields
\begin{eqnarray}
	\left. \frac{\partial I}{\partial \alpha}\right|_{\alpha=0}
	& =  & - \int_{x_L}^{x_R}  
	\frac{W_0 W_h}
	{\sqrt{E_n-W_0^2}}
	\quad{\rm d}x 
	\nonumber \\
	\nonumber\\
	&=& - \int_{x_L}^{x_R}  
	\frac{\left( \frac{1}{2}x-\frac{\ell}{x}\right) 
		\left(\frac{2 x }{ x^2 + 2\ell -1}-\frac{2  x }{ x^2+2\ell+1}\right) }
	{\sqrt{2n -  \left( \frac{1}{4} x^2 + \frac{\ell^2}{x^2} - \ell\right) 
		}}
		\quad{\rm d}x 
		\nonumber \\
		\nonumber\\
		&=& - 2\int_{u_L}^{u_R}  
		\frac{u-2{\ell} 
		}
		{\left( \left( u+2\ell\right)^2 -1\right) \sqrt{4\left( 2n+\ell\right) u -\left(  u^2 + 4{\ell^2}  \right) }}
		~{\rm d}u \label{Integral1},
	\end{eqnarray}
	where we have made a change of variable $u=x^2$.
	The values of  $u$ at the turning points are given by
	\begin{equation}
		u_{\left( \frac L R\right) 
		} = 2(2n+\ell)\mp 4\sqrt{n(n+\ell)}~.
	\end{equation}
	
\begin{figure}[htb]
		\centering
		\includegraphics[width=0.83\linewidth]{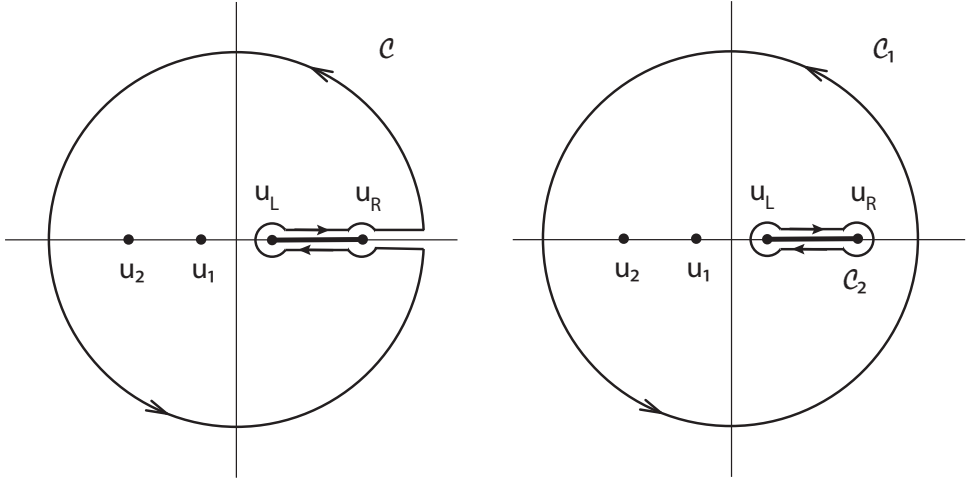}
		\caption{Complex plane calculation}
		\label{fig:poles}
	\end{figure}

	To compute the integral of (\ref{Integral1}), we embed the $u$-axis in a complex plane. It has two poles: $u_1 = -2\ell +1$ and $u_2 = -2\ell -1$ and a branch cut from $u_L$ to $u_R$. A contour integration along  $\cal C$ that includes both poles, breaks up into two contour integrals along  ${\cal C}_1$ and  ${\cal C}_2$, as illustrated in Figure \ref{fig:poles}. From these, we get
	\begin{equation}
		\int_{{\cal C}_1}+\int_{{\cal C}_2} = 2\pi i\, \left[ \mbox{Sum of residues at poles at $u_1$ and $u_2$}\right] ~.
	\end{equation}
	We find that as the radius increases, the integral $\int_{{\cal C}_1}$ goes to zero, and the integral $\int_{{\cal C}_2}$ reproduces twice the integral of (\ref{Integral1}), to produce the final result
		
	\begin{eqnarray}
		\left. \frac{\partial I}{\partial \alpha}\right|_{\alpha=0}&=& - 2\int_{u_L}^{u_R}  
		\frac{u-2{\ell} 
		}
		{\left( \left( u+2\ell\right)^2 -1\right) \sqrt{4\left( 2n+\ell\right) u -\left(  u^2 + 4{\ell^2}  \right) }}
		\quad{\rm d}u 
		\nonumber \\
		\nonumber\\
		&=& \pi\left[ \frac{4\ell -1}{\sqrt{ \left(u_L+2\ell-1\right) \left(u_R+2\ell-1\right)}} -
		\frac{4\ell +1}{\sqrt{ \left(u_L+2\ell+1\right) \left(u_R+2\ell+1\right)}}
		\right]\nonumber\\ \label{Derivative@theOrigin}
	\end{eqnarray}
where $u_{L} = 2(2n+\ell)- 4\sqrt{n(n+\ell)}$ and $u_{R} = 2(2n+\ell) + 4\sqrt{n(n+\ell)}$. The content of the square bracket is positive, and therefore the derivative is necessarily positive at $\alpha=0$. Since $\partial I/\partial \alpha\neq0$ at $\alpha=0$, $I(n,\ell,\alpha)$ will depart from the exact solution $I(n,\ell,0)=n\pi$ as $\alpha$ is increased from zero. Therefore, although the SWKB approximation is exact for $\alpha=0$, it is not exact for all $\alpha.$ In the following sections, we will increase $\alpha$ from zero to one with the goal of investigating whether the approximation is exact for the particular value $\alpha=1$ corresponding to the extended SI potential.
	
\subsection{SWKB approximation for general $\alpha$.}	

We want to analyze the behavior of the SWKB approximation for  $W(x,\ell,\alpha )=W_0+\alpha W_h$ as $\alpha$ increases from zero to the case $\alpha=1$ corresponding to the shape-invariant extended potential. 
	
We begin by finding the turning points $x_L$ and $x_R$ by factoring $E_n=W\left(x,\ell,\alpha\right)^2=0$; they are given by the solutions to $\sqrt{E_n}\pm W\left(x,\ell,\alpha\right)=0.$ For the superpotential under consideration, [\ref{fullW}], this yields a sixth-order polynomial in $x$ which in general cannot be  solved algebraically, except for particular values of $\alpha$, such as the case $\alpha=0$. Hence we solve it numerically for given $n$, $\ell$, and $\alpha$ and retain the  two real, non-negative solutions. Of these, the smaller of the two will be $x_L$ and the larger will be $x_R.$ Below, we  illustrate the square of the extended superpotential (using $\alpha=1$) for $\ell=2,$ together with the lowest three energy levels, showing the turning points for each.

\begin{figure}[!htbp]
	\centering
	\includegraphics[width=0.6\linewidth]{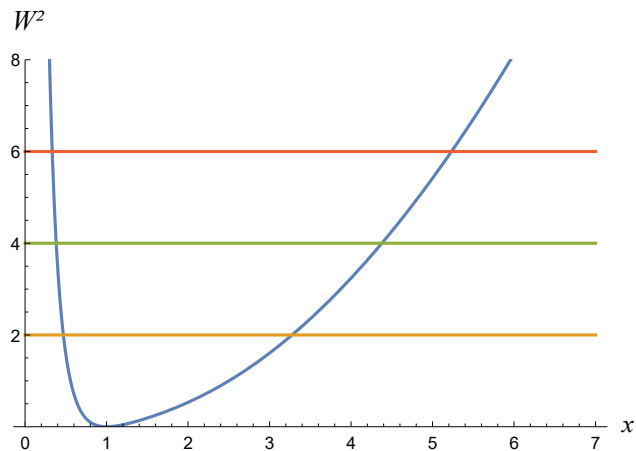}
	\caption{(Color online)The square of the extended superpotential $W^2$ for $\alpha=1$ and $\ell=1$ as a function of $x$ (blue curve), along with the first three energy levels $E_1$ (yellow), $E_2$ (green) and $E_3$ (red). The intersection of $W^2$ with the energy level $E_n$ will give the turning points for the integral corresponding to a given $n$.}
	\label{fig:turningpoints}
\end{figure}
Armed with these limits for the integral, we numerically evaluate $I(n,\ell,\alpha)$ to compare to the predictions of the SWKB approximation for various $\alpha$.

\subsubsection{Testing the numerical approximation for small $\alpha$}	
\label{sec:derivative}
	
To begin with, we will compare our numerical method to the analytical result we found for $\partial I/\partial \alpha$ in the vicinity of $\alpha=0$ as given in (\ref{Derivative@theOrigin}). To do so,we take $n$ and $\ell$ as fixed and consider $I$ as a function only of $\alpha$, then we expand $I(\alpha)$ in powers of $\alpha$ about the point $\alpha=0$. 
	
For small $\Delta\alpha$, 
	\begin{equation}
	I(\Delta\alpha)=I(0) + \left. \frac{\partial I}{\partial \alpha}\right|_{\alpha=0} \Delta \alpha + \left.\frac{1}{2}\frac{\partial^2 I}{\partial \alpha^2}\right|_{\alpha=0} \Delta \alpha^2+ \mathcal{O}(\Delta\alpha^3),
	\end{equation}
which, for simplicity of notation, we write as
	\begin{equation}
	I(\Delta\alpha)=I(0) + I'(0)\Delta\alpha+\frac{1}{2}I''(0)\Delta\alpha^2+\mathcal{O}(\Delta\alpha^3).
	\end{equation}
	
Therefore, by numerically finding $I(\Delta\alpha)$ for small $\Delta\alpha$, we can find a numerical approximation for the slope at zero:
	\begin{equation}
		I_{num}'(0)= \frac{I(\Delta\alpha)-I(0)}{\Delta \alpha},		
	\end{equation} 
	 where the difference between the actual and the numerical slopes at zero is given by
	 \begin{equation}
	 	I'(0) - I_{num}'(0)= -\frac{1}{2}I''(0)\Delta\alpha+\mathcal{O}(\Delta\alpha^2).\label{eq:numderiv}
 	\end{equation} 
	 
Equation (\ref{Derivative@theOrigin}) gives an exact value for the derivative $I'(0)$ for a given $n$ and $\ell$.  For instance, in the case $n=1, \ell=1$: $I'(0)=\left( 3/\sqrt{17}-5/7 \right)\pi \approx 0.0418497$. Table \ref{table:derivatives} shows  a decimal approximation of the analytical values in comparison to the numerical values of the slope at $\alpha=0$ for various values of $n$ and $\ell$, using $\Delta\alpha=10^{-5}$.

\Table{\label{table:derivatives} 	Analytical derivative $I'(0)$, compared to an approximation $I'_{num}(0)$ based on our numerical approximation and an expansion in small $\alpha$, for the case $\alpha=10^{-5}$. The ratio $I'_{num}(0)/I'(0)$ shows strong agreement between our numerical methods and the exact result.}

\br
			n	&$\ell$	&Analytical  $I'(0)$ 	&Numerical $I'_{num}(0)$ &Numerical/Analytical \\
 && from Eq.(\ref{Derivative@theOrigin})   & 	 &$I'_{num}(0)/I'(0)$\\
\mr
			1	&1	&0.0418497 & 0.0418486 &0.999976\\
\ms
			2	&1	&0.0464776& 0.0464766& 0.999980\\
		\ms
			2	&2	&0.00453316 &0.00453305& 0.999975 \\
			\ms
			3	&1	&  0.0457412 & 0.0457404& 0.999982\\
		\ms
			3	&2	& 0.00487758 &0.00487747&0.999978 \\
			\ms
			3	&3	& 0.00131056 &0.00131053 & 0.999975\\
		\ms
			4	&1	& 0.0439064  &0.0439057& 0.999984\\
		\ms
			4	&2	& 0.00495553 &0.00495543	&0.999980 \\
		\ms
			4	&3	& 0.00138774 &0.00138771 &0.999977 \\
		\ms
			4	&4	&  0.00054823 &0.00054822 &0.999975 \\
			\ms
			4	&10	&   0.0000237426 	&0.0000237418	& 0.999969\\
			\ms
			4	&100	& 3.70267E-9 &3.70253E-9 	&0.999961 \\
		\ms
			4	&1000	& 3.90355E-13	&3.90339E-13 & 0.999960\\
			\ms
			1000 &1000	&  3.47100E-11	&3.47092E-11 	&0.999975\\
		
\br
\endTable

As predicted, the derivative is positive in all cases at $\alpha=0$ for both the numerical and analytical calculations. Additionally, for $\Delta\alpha=10^{-5},$ the numerical approximation agrees with the analytical solution to an accuracy of $>99.99\%$ for all value of $n$ and $\ell$ tested. We can check the convergence of our numerical solution by manipulating Eq.(\ref{eq:numderiv}) to obtain the fractional difference:
\begin{equation}
\Gamma\equiv\frac{I'(0)-I'_{num}(0)}{I'(0)}=\frac{-I''(0)}{2I'(0)}\Delta\alpha + \mathcal{O}(\Delta\alpha^2).	
\end{equation}	
Therefore, the closeness of $\Gamma$ to zero gives a measurement of the accuracy of our numerical approximation.
 
Since we want to examine small $\Delta\alpha$, we write it as: $\Delta\alpha=10^{-\lambda}$, for some positive value $\lambda$. Ignoring terms of second-order and higher in $\Delta\alpha$ and taking the logarithm of both sides yields:
\begin{equation}
{\log_{10}}\, (\Gamma) = -\lambda + {\log_{10}} \left(\frac{I''(0)}{2I'(0)}\right).	
\end{equation}

\begin{figure}[h]
		\centering
		\includegraphics[width=0.83\linewidth]{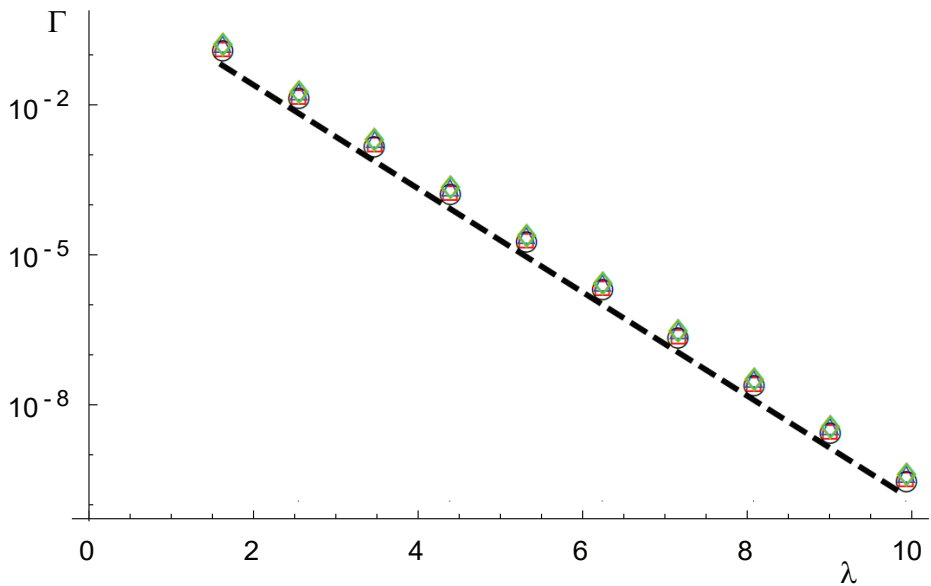}
		\caption{(Color online) Fractional difference $\Gamma$ as a function of exponent $\lambda$ for the following cases:
			$n=1,\ell=1$: red squares, $n=1,\ell=2$:  blue triangles, $n=2,\ell=1$: black circles, $n=1000,\ell=1000$, green diamonds. The dashed line is a line of slope -1 on this logarithmic scale to guide the eye. }
		\label{fig:convergence}
	\end{figure}

Plotting ${\log_{10}}\, (\Gamma)$ vs. $\lambda$ therefore should give a straight line with slope -1 for all values of $n$ and $\ell$. The y-intercept of the graph should be given by ${\log_{10}} \left (\frac{I''(0)}{2I'(0)} \right)$, which should vary with $n$ and $\ell$; however, if the ratio of derivatives $I''(0)/I'(0)$ does not vary by orders of magnitude, the y-intercept should vary only weakly with $n$ and $\ell$ due to the logarithm. We investigate this dependence in Figure~\ref{fig:convergence}. 

As expected, these graphs appear to be linear on this semi-log scale in $\lambda$, 
with only weakly varying intercept for each of the values of $n$ and $\ell$ tested. 
The slope of each graph is consistent with our prediction of a slope of -1. This verifies that our approximation is converging as predicted with a deviation from the analytical value that is linear in $\Delta\alpha$. 

\subsubsection{Validity of the SWKB approximation as a function of $\alpha$}	
\begin{figure}[h!]
		\centering
		\includegraphics[width=0.83\linewidth]{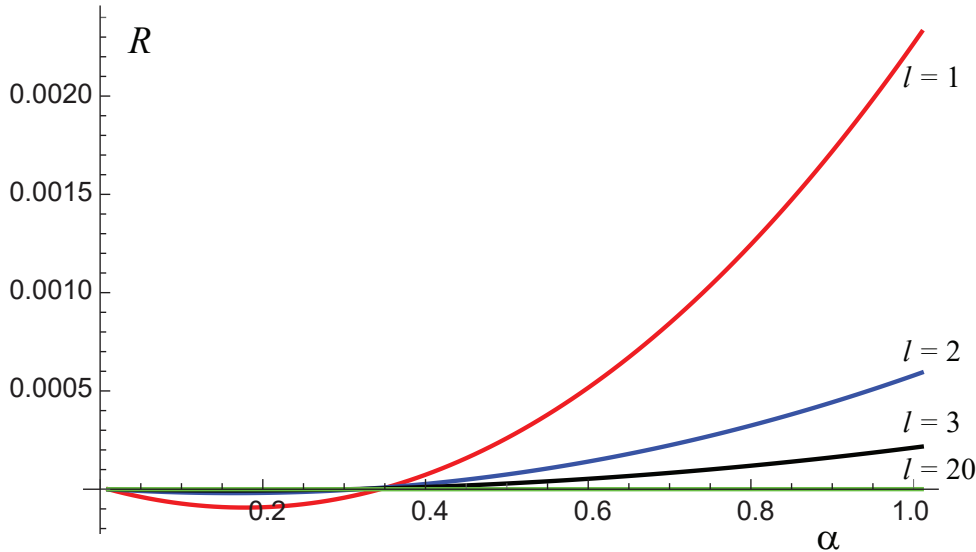}
		\caption{(Color online) Residual $R$ as a function of $\alpha$ for $n=1$ in the following cases:
			$\ell=1$: red, $\ell=2$: blue, $\ell=3$: black, and $\ell=20$: green. 
		}
		\label{fig:varyingalpha}
	\end{figure}

Having tested the robustness of the numerical approach,
we will now investigate the validity of the SWKB approximation
	$I(n,\ell,\alpha)\approx n\pi $ to see how its accuracy varies as $\alpha$ varies in the range $0\leq\alpha\leq 1$. 
To do so, we define the residual quantity 
\begin{equation}
R=1 -\frac{I(n,\ell,\alpha)}{n\pi}.
\end{equation}
Note that $R=0$ in the case in which the approximation is exact; in general, smaller $R$ will correspond to a more precise approximation, while a large value of $R$ will correspond to a less precise approximation. In Figure~\ref{fig:varyingalpha}, we have have plotted $R$ for  $n=1$ and various values of $\ell$.

For all values of $n$ and $\ell$, the function $I(\alpha )$ is equal to $n\pi$ at $\alpha =0$ as predicted by SWKB with an initial negative slope in $R$ as predicted in Sec.~\ref{sec:derivative}, since $ \frac{dR}{d\alpha} = -\frac1{n\pi} \, \frac{dI}{d\alpha}$. For each of the values of $n$ and $\ell$ examined, the integral crosses $n\pi$ at some later value of $\alpha$, leading to a positive value of $R$ for $\alpha=1$ in each case.

\begin{figure}[h!]
	\centering
	\includegraphics[width=0.75\linewidth]{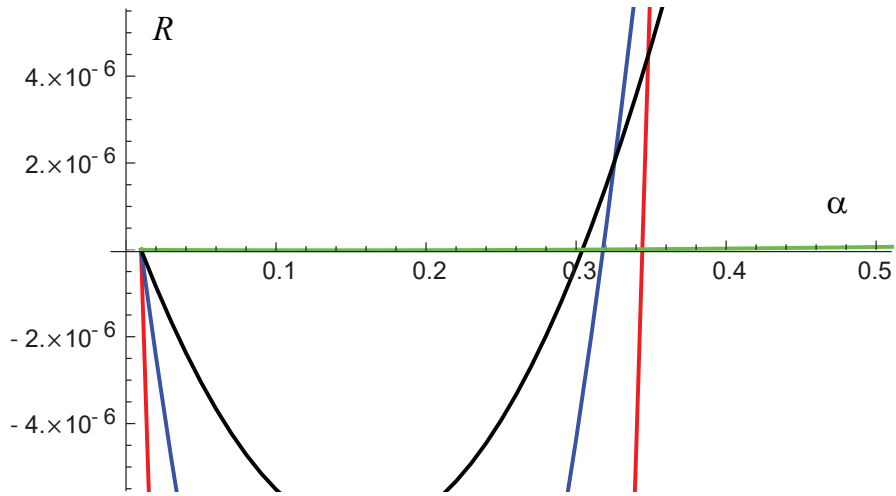}
	\caption{(Color online) Magnification of the region around $\alpha=0.3$ for Figure \ref{fig:varyingalpha}.
	}
	\label{fig:varyingalphazoom}
\end{figure}

From this graph, it is not clear whether the zero of $R$ for positive $\alpha$ occurs at the same value of $\alpha$ for different $n$ and $\ell$. Therefore, we zoom in on this zero and discover that it does, in fact, vary with the parameters as seen in Figure~\ref{fig:varyingalphazoom}.

	\subsubsection{Inexactness of the SWKB approximation for the extended superpotential}	

We therefore see that the SWKB approximation is not, in fact, exact for the value $\alpha=1$ corresponding to the extended shape-invariant superpotential. As a final step, we examine how this deviation from SWKB varies with $n$ and $\ell$. Returning to the extended superpotential, we see that in the limit of large $\ell$, the $\hbar$-dependent extension $W_h$ to the  superpotential becomes negligible compared to $W_0$ ({\it cf} Eqs.~(\ref{eq:W0},\ref{eq:hbar1W})). Therefore, in this limit, we should approach the exact condition for SWKB. Additionally, the validity of the SWKB approximation increases for large $n$, so $R$ should also approach zero for large $n$.

\begin{figure}[htb]
	\centering
	\includegraphics[width=0.9\linewidth]{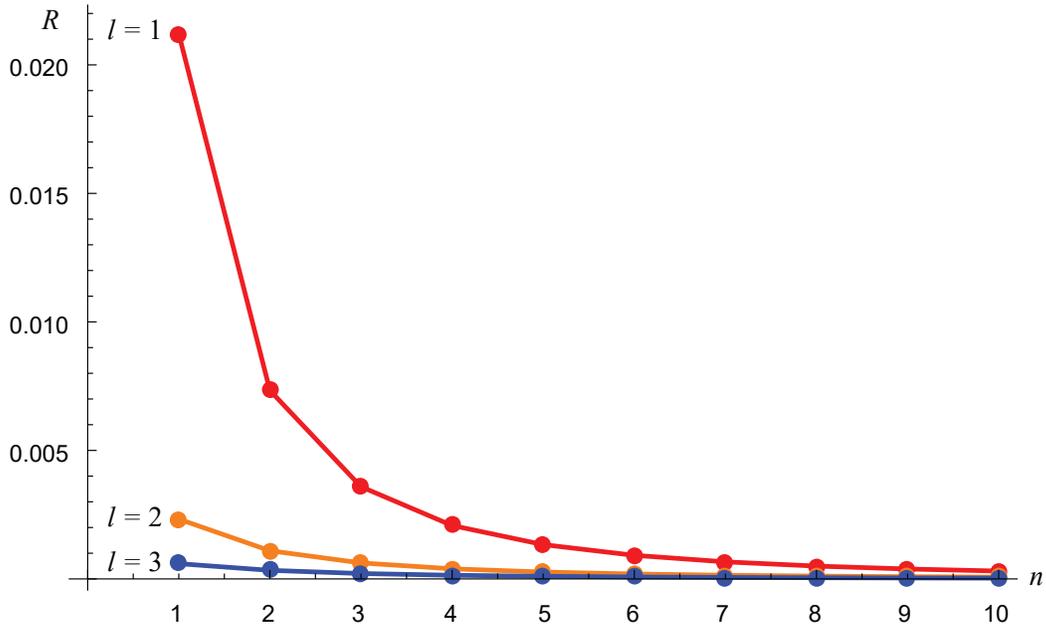}
	\caption{(Color online) $R$ as a function of $n$ for the following cases: $\ell=1$ (red), $\ell=2$ (orange), and $\ell=3$ (blue).
	}
	\label{fig:RES}
\end{figure}

We plot $R$  as a function of $n$ for $\alpha=1$ and various $\ell$ in Figure~\ref{fig:RES}. We see that, as expected,  $R$ does indeed decrease with increasing $n$ and $\ell$.

\begin{figure}[htbp]
	\centering
	\includegraphics[width=0.9\linewidth]{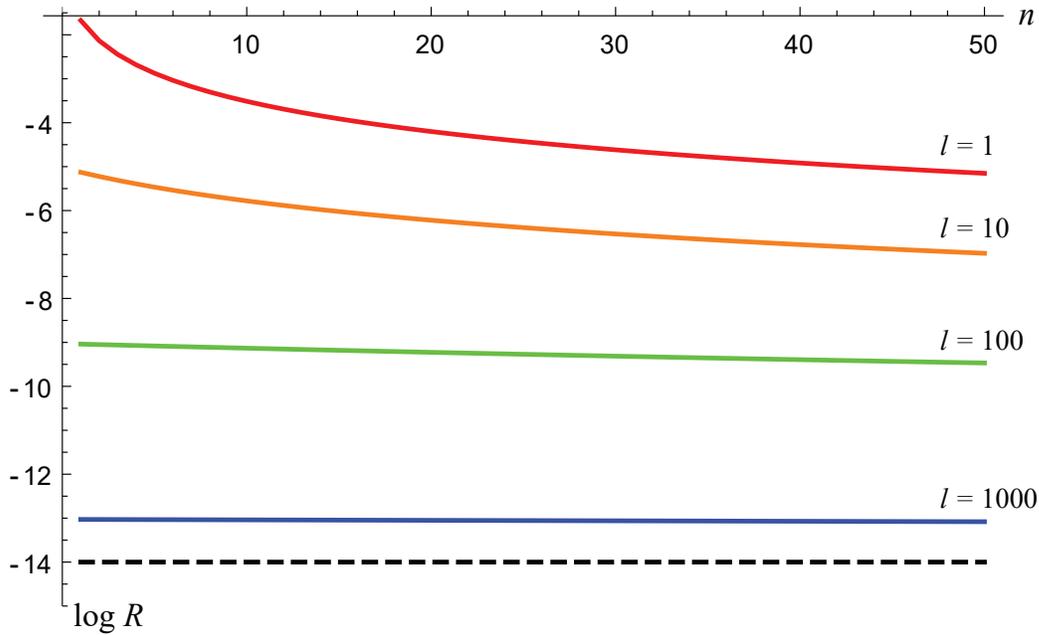}
	\caption{(Color online) The residual $\log_{10} R$ as a function of $n$ for the following cases: $\ell=1$ (red), $\ell=10$ (orange), $\ell=100$ (green), and $\ell=1000$ (blue). The guiding dashed line shows the constant value $10^{-14}$ to illustrate the proximity of $R$ to zero for $\ell=1000$, for any value of $n$.}
	
	\label{fig:LogRES}
\end{figure}
	
To examine this convergence towards the exact solution, and to look at larger values of $\ell$, we view this data on a logarithmic plot, and see that after an initial curvature near $n=1$, the curves appear nearly linear and parallel.

\section{Conclusions}

Conventional shape-invariant superpotentials have the important property of having exactly solvable spectra. These superpotentials also share the property of making the SWKB quantization condition exact.

Extended SI superpotentials are exactly solvable due to the properties of additive shape-invariance that they share with the conventional superpotentials. However, we have shown that additive shape-invariance does not guarantee SWKB exactness by presenting a counterexample: the extended radial oscillator. This result suggests that the exactness of the conventional superpotentials may be connected to their $\hbar$-independence and may suggest that further investigation of the role of $\hbar$-dependence of the SWKB approximation could play a role in better understanding the properties of these extensions. 

\pagebreak


\begin{thebibliography}{10}
	\expandafter\ifx\csname url\endcsname\relax
	\def\url#1{\texttt{#1}}\fi
	\expandafter\ifx\csname urlprefix\endcsname\relax\def\urlprefix{URL }\fi
	\expandafter\ifx\csname href\endcsname\relax
	\def\href#1#2{#2} \def\path#1{#1}\fi
	
	\bibitem{Jeffreys}
	H.~Jeffreys, On certain approximate solutions of linear differential equations
	of the second order, Proc. of the London Math. Soc. 23 (1924) 428--436.
	
	\bibitem{Wentzel}
	G.~Wentzel, Eine verallgemeinerung der quantenbedingungen f{\"u}r die zwecke
	der wellenmechanik, Zeitschrift f{\"u}r Physik 38 (1926) 518--529.
	
	\bibitem{Kramers}
	H.~A. Kramers, Wellenmechanik und halbz{\"a}hlige quantisierung, Zeitschrift
	f{\"u}r Physik 39 (1926) 828--840.
	
	\bibitem{Brillouin}
	L.~Brillouin, La m{\'e}canique ondulatoire de schr{\"o}dinger: une m{\'e}thode
	g{\'e}n{\'e}rale de resolution par approximations successives, Comptes Rendus
	de l'Academie des Sciences 183 (1926) 24--26.
	
	\bibitem{Bender_Orzag}
	C.~M. Bender, S.~A. Orzag, Advanced Mathematical Methods for Scientists and
	Engineers, McGraw-Hill, New York, 1978, contains a detailed and modern
	description.
	
	\bibitem{Bender}
	C.~M. Bender, K.~Olaussen, P.~Wang, Numerological analysis of the {WKB}
	approximation in large order, Phys. Rev. D 16 (1977) 1740--1748.
	
	\bibitem{Comtet}
	A.~Comtet, A.~D. Bandrauk, D.~K. Campbell, Exactness of semiclassical bound
	state energies for supersymmetric quantum mechanics, Phys. Lett. B 150 (1985)
	159--162.
	
	\bibitem{Infeld}
	L.~Infeld, T.~E. Hull, The factorization method, Rev. Mod. Phys. 23 (1951)
	21--68.
	
	\bibitem{Dutt_SUSY}
	R.~Dutt, A.~Khare, U.~Sukhatme, Supersymmetry, shape invariance and exactly
	solvable potentials, Am. J. Phys 56 (1988) 163--168.
	
	\bibitem{Dutt}
	R.~Dutt, A.~Khare, U.~Sukhatme, Exactness of supersymmetry {WKB} spectra for
	shape-invariant potentials, Phys. Lett. B 181 (1986) 295--298.
	
	\bibitem{Adhikari}
	R.~Adhikari, R.~Dutt, A.~Khare, U.~P. Sukhatme, Higher-order {WKB}
	approximations in supersymmetric quantum mechanics, Phys. Rev. A 38 (1988)
	1679.
	
	\bibitem{Raghunathan}
	K.~Raghunathan, M.~Seetharaman, S.~S. Vasan, On the exactness of the {SUSY}
	semiclassical quantization rule, Phys. Lett. B 188 (1987) 351--352.
	
	\bibitem{Barclay}
	D.~T. Barclay, C.~J. Maxwell,
	\href{http://www.sciencedirect.com/science/article/pii/037596019190869A}{Shape
		invariance and the {SWKB} series}, Phys. Lett. A 157~(6) (1991) 357 -- 360.
	\newblock \href
	{http://dx.doi.org/https://doi.org/10.1016/0375-9601(91)90869-A}
	{\path{doi:https://doi.org/10.1016/0375-9601(91)90869-A}}.
	\newline\urlprefix\url{http://www.sciencedirect.com/science/article/pii/037596019190869A}
	
	\bibitem{Yin}
	C.~Yin, C.~Zhaungqi, Q.~Shen, Why {SWKB} approximation is exact for all {SIP}s,
	Ann. Phys. (N. Y.) 325 (2010) 528--534.
	
	\bibitem{Quesne1}
	C.~Quesne, Exceptional orthogonal polynomials, exactly solvable potentials and
	supersymmetry, J. Phys. A 41 (2008) 392001.
	
	\bibitem{Quesne2}
	C.~Quesne, Solvable rational potentials and exceptional orthogonal polynomials
	in supersymmetric quantum mechanics, Sigma 5 (2009) 084.
	
	\bibitem{Odake1}
	S.~Odake, R.~Sasaki, Infinitely many shape invariant discrete quantum
	mechanical systems and new exceptional orthogonal polynomials related to the
	{Wilson and Askey-Wilson} polynomials, Phys. Lett. B 682 (2009) 130--136.
	
	\bibitem{Odake2}
	S.~Odake, R.~Sasaki, Another set of infinitely many exceptional $(x_\ell)$
	{Laguerre} polynomials, Phys. Lett. B 684 (2010) 173--176.
	
	\bibitem{Witten}
	E.~Witten, Dynamical breaking of supersymmetry, Nucl. Phys. B 185 (1981)
	513--554.
	
	\bibitem{Solomonson}
	P.~Solomonson, J.~W. {Van Holten}, Fermionic coordinates and supersymmetry in
	quantum mechanics, Nucl. Phys. B 196 (1982) 509--531.
	
	\bibitem{CooperFreedman}
	F.~Cooper, B.~Freedman, Aspects of supersymmetric quantum mechanics, Ann. Phys.
	146 (1983) 262--288.
	
	\bibitem{Cooper-Khare-Sukhatme}
	F.~Cooper, A.~Khare, U.~Sukhatme, Supersymmetry in Quantum Mechanics, World
	Scientific, Singapore, 2001.
	
	\bibitem{Gangopadhyaya-Mallow-Rasinariu}
	A.~Gangopadhyaya, J.~Mallow, C.~Rasinariu, Supersymmetric Quantum Mechanics: An
	Introduction, World Scientific, Singapore, 2010.
	
	\bibitem{Miller}
	W.~Miller, Lie Theory and Special Functions, Academic Press, New York, 1968.
	
	\bibitem{gendenshtein1}
	L.~E. Gendenshtein, Derivation of exact spectra of the schrodinger equation by
	means of supersymmetry, JETP Lett. 38 (1983) 356--359.
	
	\bibitem{gendenshtein2}
	L.~E. Gendenshtein, I.~V. Krive, Supersymmetry in quantum mechanics, Sov. Phys.
	Usp. 28 (1985) 645--666.
	
	\bibitem{CGK}
	F.~Cooper, J.~N. Ginocchio, A.~Khare, Relationship between supersymmetry and
	solvable potentials, Phys. Rev. D 36 (1987) 2458--2473.
	
	\bibitem{Tanaka}
	T.~Tanaka, N-fold supersymmetry and quasi-solvability associated with
	x-2-{Laguerre} polynomials, J. Math. Phys. 51 (2010) 032101.
	
	\bibitem{Odake3}
	S.~Odake, R.~Sasaki, Exactly solvable quantum mechanics and infinite families
	of multi-indexed orthogonal polynomials, Phys. Lett. B 702 (2011) 164--170.
	
	\bibitem{Odake4}
	S.~Odake, R.~Sasaki, Extensions of solvable potentials with finitely many
	discrete eigenstates, J. Phys. A 46 (2013) 235205.
	
	\bibitem{Quesne2012a}
	C.~Quesne, Novel enlarged shape invariance property and exactly solvable
	rational extensions of the {Rosen-Morse II} and {Eckart} potentials, Sigma 8
	(2012) 080.
	
	\bibitem{Quesne2012b}
	C.~Quesne, Revisiting (quasi-)exactly solvable rational extensions of the
	{Morse} potential, Int. J. Mod. Phys. A 27 (2012) 1250073.
	
	\bibitem{Ranjani1}
	S.~S. Ranjani, P.~K. Panigrahi, A.~K. Kapoor, A.~Khare, A.~Gangopadhyaya,
	Exceptional orthogonal polynomials, {QHJ} formalism and {SWKB} quantization,
	J. Phys. A 45 (2012) 055210.
	
	\bibitem{Ranjani2}
	S.~S. Ranjani, R.~Sandhya, A.~K. Kapoor, Shape invariant rational extensions
	and potentials related to exceptional polynomials, Int. J. Mod. Phys. A 30
	(2015) 1550146.
	
	\bibitem{Bougie2010}
	J.~Bougie, A.~Gangopadhyaya, J.~V. Mallow, Generation of a complete set of
	additive shape-invariant potentials from an {Euler} equation, Phys. Rev.
	Lett. 105 (2010) 210402.
	
	\bibitem{symmetry}
	J.~Bougie, A.~Gangopadhyaya, J.~V. Mallow, C.~Rasinariu, Supersymmetric quantum
	mechanics and solvable models, Symmetry 4 (2012) 452--473.
	
	\bibitem{mahdi2016}
	K.~Mahdi, Y.~Kasri, Y.~Grandati, A.~B{\'e}rard, {SWKB} and proper quantization
	conditions for translationally shape-invariant potentials, Eur. Phys. J. Plus
	131 (2016) 259.
	
\end{thebibliography}

\end{document}